# DESTINATION INFORMATION MANAGEMENT SYSTEM FOR TOURIST

Abdulhamid Shafii Muhammad[1], Gana Usman[2]

Dept. of Cyber Security Science, Fed. Univ. of Techno. Minna, Niger State, Nigeria.[1]
Department of Physics, Bayero University Kano, Kano State, Nigeria.[2]
*shafzon@yahoo.com*[1]

*Abstract*

*The use of information and communication technology in our day to day activities is now unavoidable. In tourism developments, destination information and management systems are used to guide visitors and provide information to both visitors and management of the tour sites. In this paper, information and navigation system was designed for tourists, taking some Niger state of Nigeria tourism destinations into account. The information management system was designed using Java Applet (NetBeans IDE 6.1), Hypertext MarkUp Language (HTML), Personal Home Page (PHP), Java script and MySQL as the back-end integration database. Two different MySQL servers were used, the MySQL query browser and the WAMP5 server to compare the effectiveness of the system developed.*

**Introduction**

The proliferation of the Web over the last few years led companies and organizations to try to exploit the Web for many different activities. Tourism is one of the most important applications of e-commerce. Several major tourism actors and even the new comers (information technology companies mainly) have an established Web presence, visited by many thousands of visitors every day, offering e-commerce opportunities for business to business transactions or business to customer (tourist) transactions. One particular class of tourism applications in the Web is Destination Information Systems (DIS) or Destination Management Systems (DMS). These systems typically provide in the Web, information about the tourism offerings of a given Destination and may promote e-commerce activities to the potential visitor[1].

The existing DMS's however do not support advanced models of interaction between tourists (or prospective tourists) of a Destination, nor interaction between tourists and locals. It is believed that this is a serious limitation of the existing DMS's, and therefore this work will propose an expanded functionality that provides the tourists with intelligent interactions based on a virtual community concept of tourists and locals that has a common interest theme, "Tourism at Destination". Information systems that support interactions of a virtual community over the Web, which has some specific interests (the glue of this community), are usually called Community based Information Systems (CIS). Some of them have user populations of the order of tens of thousands who are repeatedly visiting the community site. However the support that the existing CIS's offer is of general purpose and they cannot be easily used to offer advanced functionality for tourism related communities.

It is considered that it is very important both for tourists and for Destinations to support advanced information models enabling the interaction of tourists and locals for tourism related subjects. Such systems will bridge the "Community Gap", which is the lack of interaction among tourists and locals at a particular Destination.

**Problem Description.** Tourists world over are always ready to explore new destinations, but the problem is that most at times they end up being disappointed after visiting some sites. This is as a result of misinformation and lack of planning by the site managers most at times.

- A tourist needs to have full information on what he/she should expect at the destination.
- A tourist needs to have a tour guide that will help him/she to navigate around the destinations of choice.

**Literature Review**

There is a mobile system named *Minotaurus* which combines the above technologies to achieve the development of a mobile, multimedia tourist information system [2]. This system, consisting of a portable computer (Laptop), having large storage capacities, capabilities of wireless connection to a worldwide information network and provide to their users many functionalities like access to WWW, shopping, banking, reservations and other transactions.

A model supporting intelligent interactions of tourists with other tourists and locals and the tourism information of a particular destination before, during and after the trip. The approach tries to bridge the "Community Gap" which is the lack of interactions among tourists and between tourists and locals at a particular destination. Community interactions are very important both for prospective visitors and for destinations for many reasons including, greater independence and self-planning in the visit's design, exploitation of the local society knowledge about the destination, as well as promotion of regional policies and collective purchases of services from prospective visitors. Modern information technology has become ubiquitous, supporting visitors with a variety of devices ranging from handy devices, to community walls, to paper interfaces, to home PCs [3].

Information systems that support interactions of a virtual community over the Web, which has some specific interests (the glue of this community), are usually called Community based Information Systems (CIS). Existing CIS's in the Web focus to foster social objectives like building community cohesion, enhancing community awareness in local decision making, developing economic opportunities in disadvantaged communities, and enhanced training [4]. Some of them have user populations of the order of tens of thousands who are repeatedly visiting the community site. However the support that the existing CIS's offer is of general purpose and they cannot be easily used to offer advanced functionality for tourism related communities.

In a similar research carried out by [5], Nigeria's determined efforts to promote tourism industry since 1991 were enumerated and Lagos state was considered as a case study. This includes the establishment of the National Policy on Tourism and the National Tourism Development Corporation (NTDC) with the objective of making Nigeria the ultimate tourism destination in Africa. Tourism spatial and attribute data gathered were classified into three categories – Cultural, Ecological and Modern day tourism. A relational GIS database was created using Arc View and graphic (map), picture and sound data were integrated into the multimedia GIS database. The various software which made this possible were; Microsoft Excel, Arcview, ArcGIS, AutoCAD release 14, CAD Overlay, Media Studio Pro 5, Video Edition and ULEAD Video Studio with a firewire adapter. The various outputs from the database include a spatial queries, analogue and electronic Tourism Atlas, Encyclopedia, a Digital library of Tourism[6],[7].

The research work carried out by [8] reviled that even for the experienced information professional, designing an efficient multipurpose information access structure can be a very difficult task. The foundations for building mobile environmental information systems (MEIS) that require a multidisciplinary approach. MEIS require expertise from environmental biology, geography and mobile technology[9].

**Methodology**

System analysis is the process of examining an existing system in order to modify the existing system or design a new system entirely. System analysis is carried out to achieve mainly two aims namely:

1. To have a clear understanding of the system or the process. This will help in the design of a new system.
2. Analysing the system will bring about identifying its problem and hence knowing the reason for its inefficiency.

The Unified Modeling Language (UML) is a family of graphical notations, backed by single meta-model, that help in describing and designing software systems, particularly software systems built using the object-oriented style . That is a somewhat simplified definition . In fact, the UML is a few different things to different people . This comes both from its own history and from the different views that people have about what makes an effective software engineering process. The UML was used to first design the proposed system. The Use-Case diagram and the Class diagram are presented below.

The Use-Case Model captures the requirements of a system. Use-cases are a means of communicating with users and other stakeholders about what the system is intended to do. A Use-Case Diagram shows the interaction between the system and entities external to the system. These external entities are referred to as Actors. Actors represent roles which may include human users, external hardware or other systems. The tourist is the actor in this case.

The Use-Case Diagram For the Proposed System

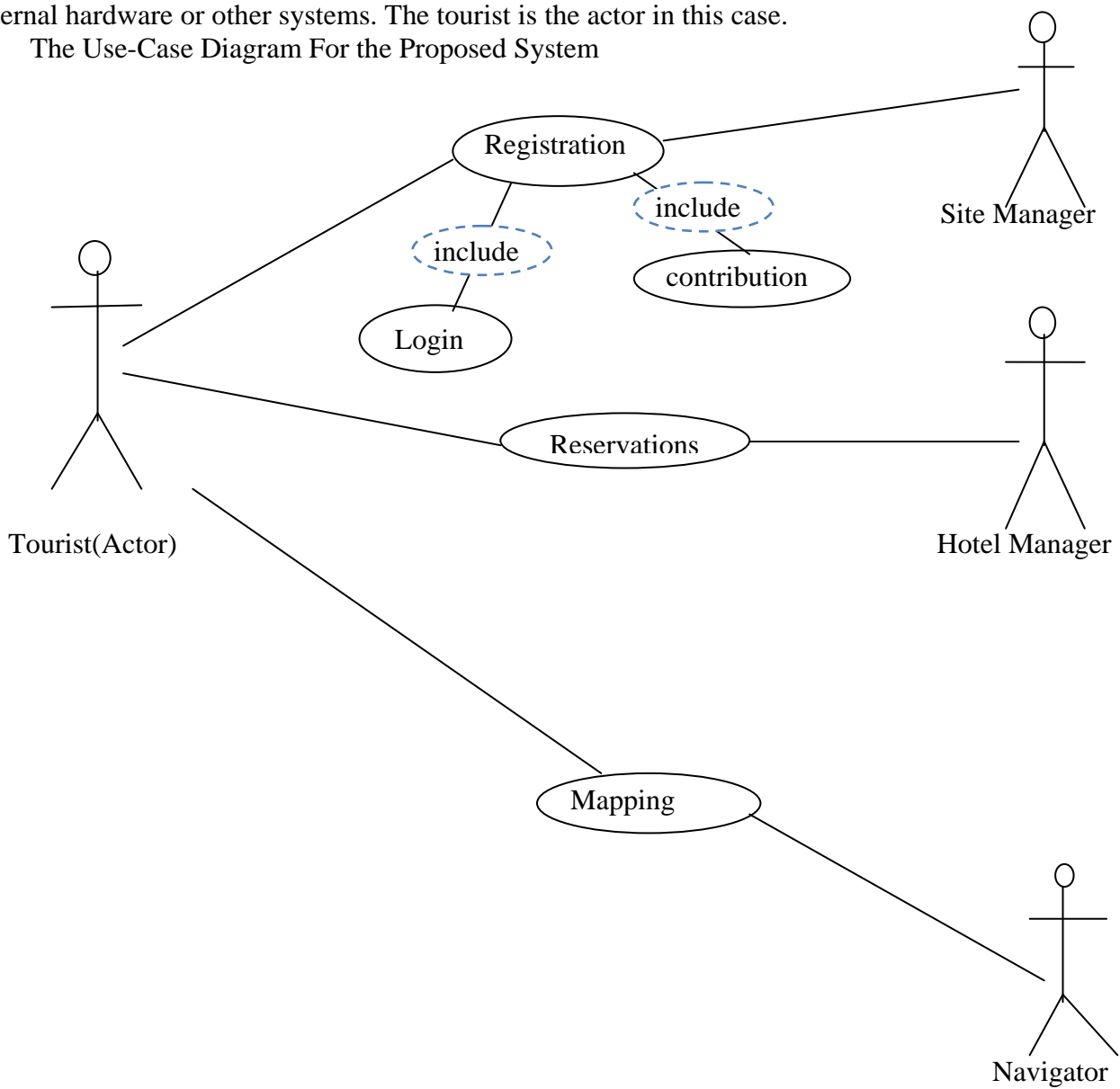

**Figure 1:** Use-Case Diagram for Destination Information Management System

Class Diagram For the Proposed System

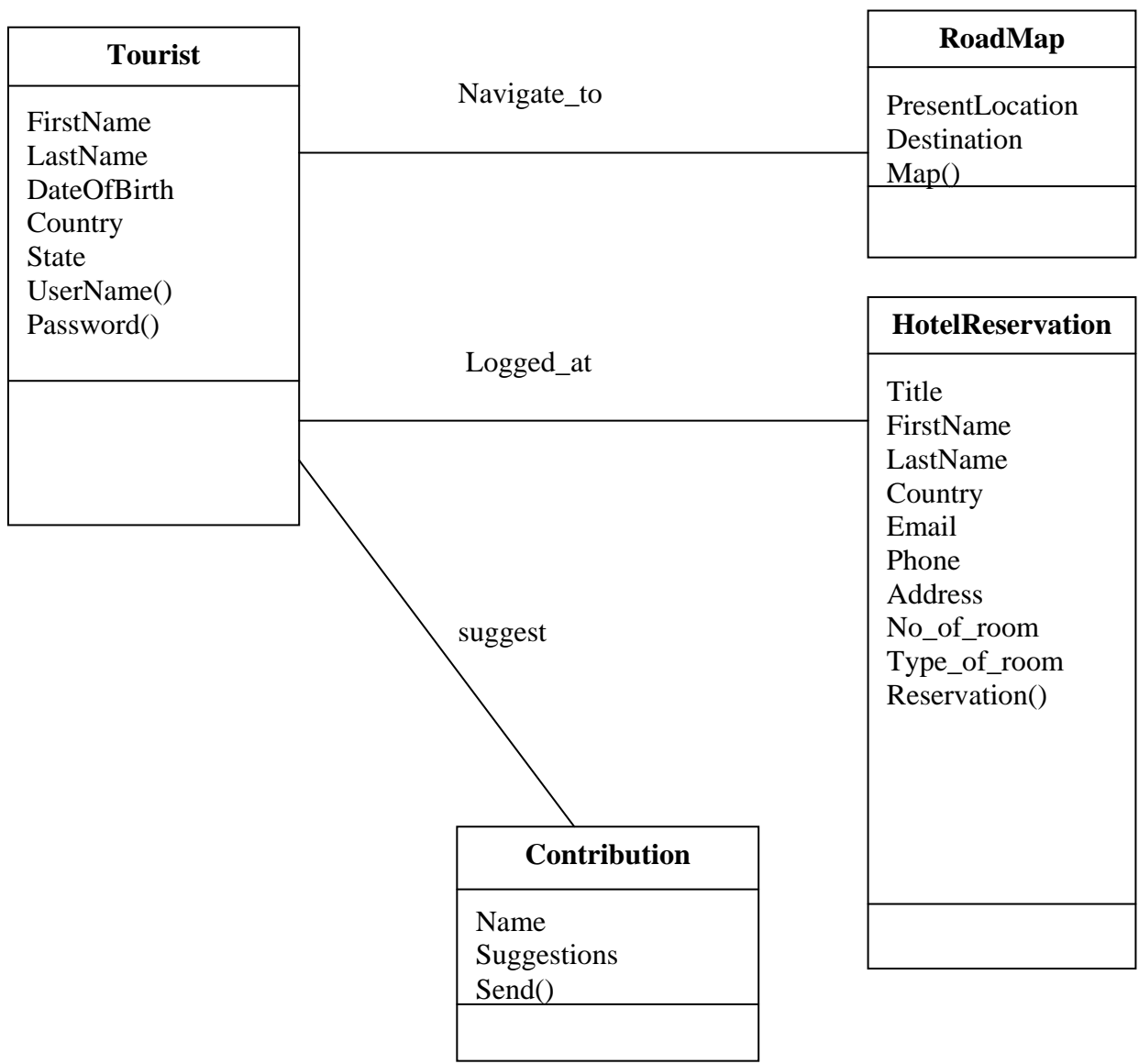

**Figure 2:** Class Diagram for Destination Information Management System

**Implementation**

The programming languages used in this work are Java Applet, PHP and HTML. With MySQL as the back-end integration database. The choice of these programming languages is based on the features of the languages that makes them more appropriate for this work.

*Starting the Program*

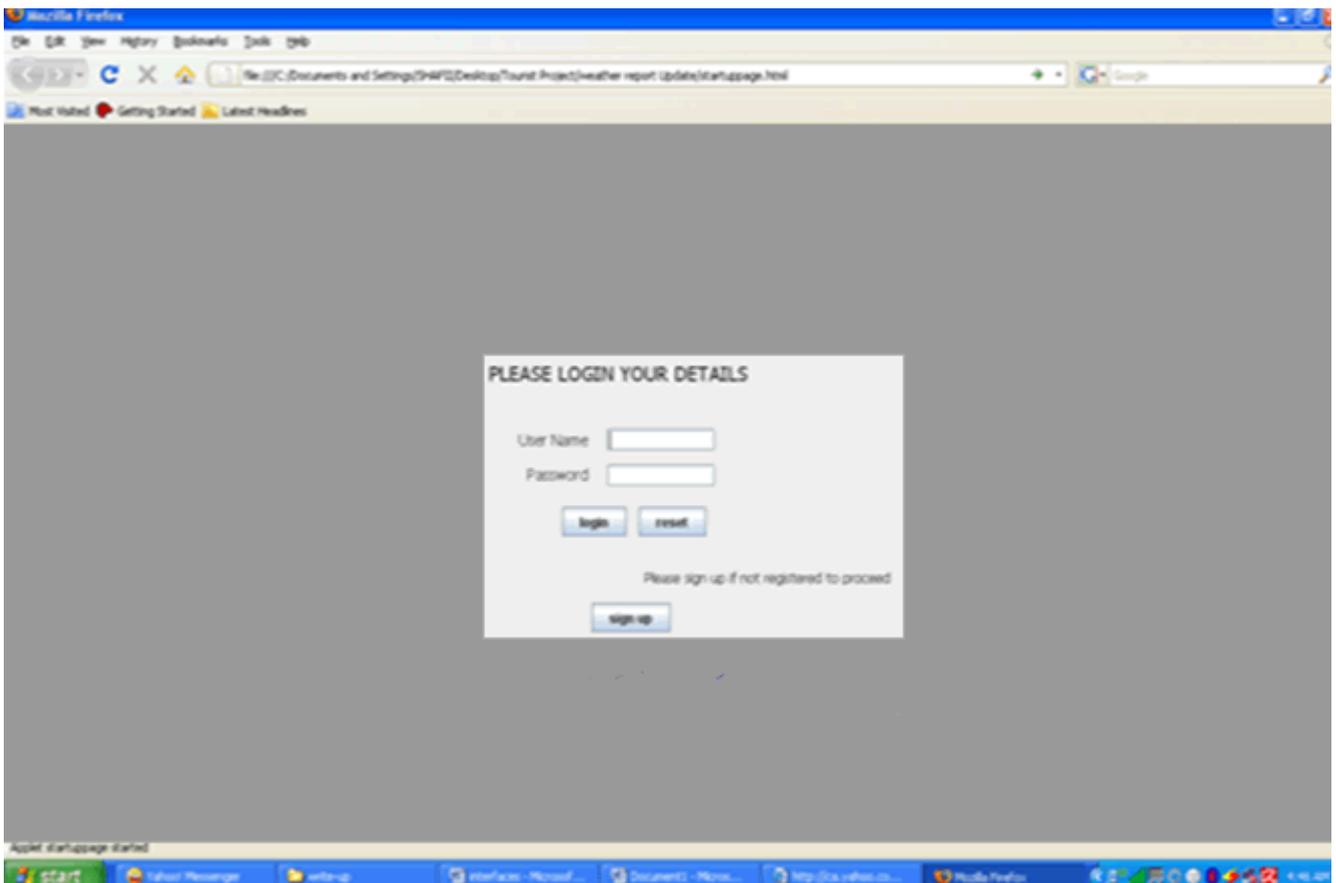

Figure 3: Login Page

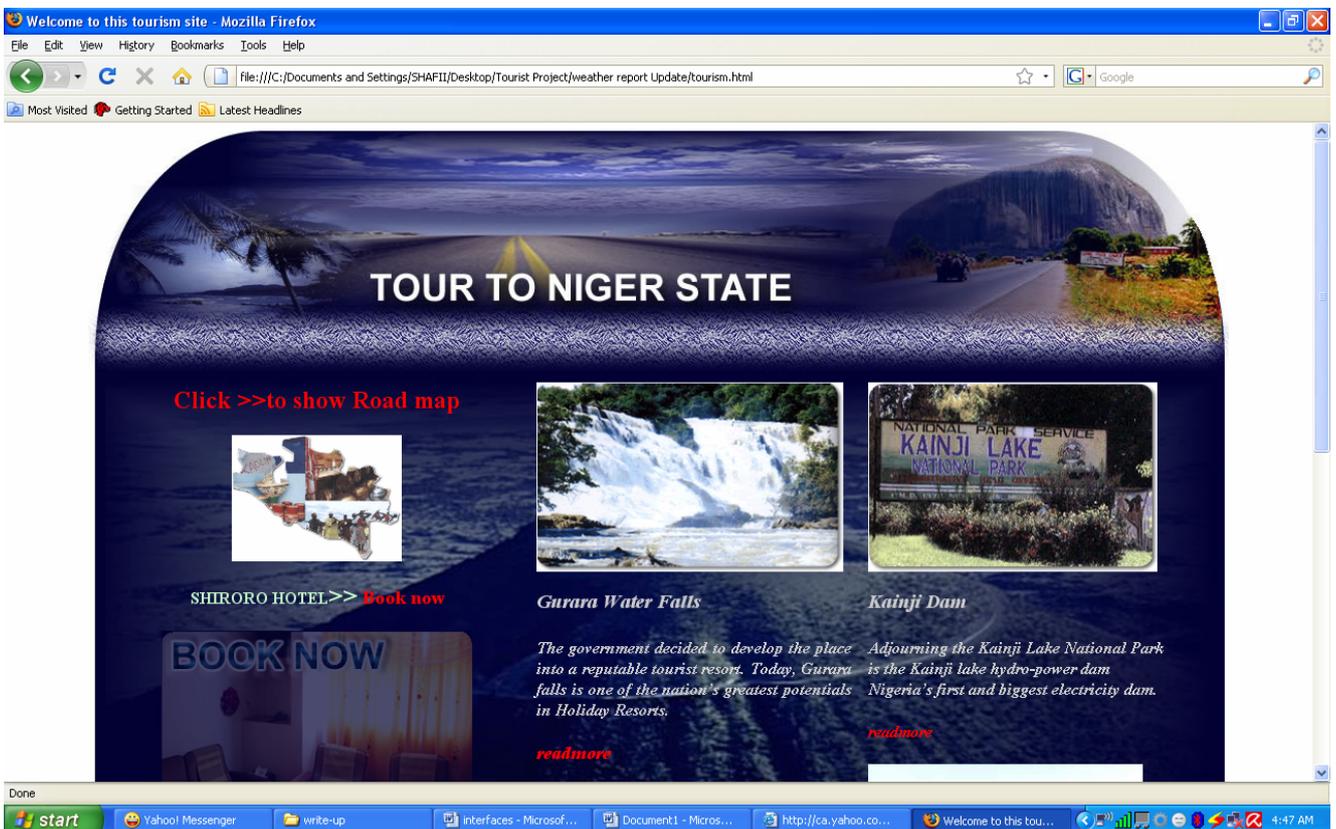

Figure 4: Main Menu

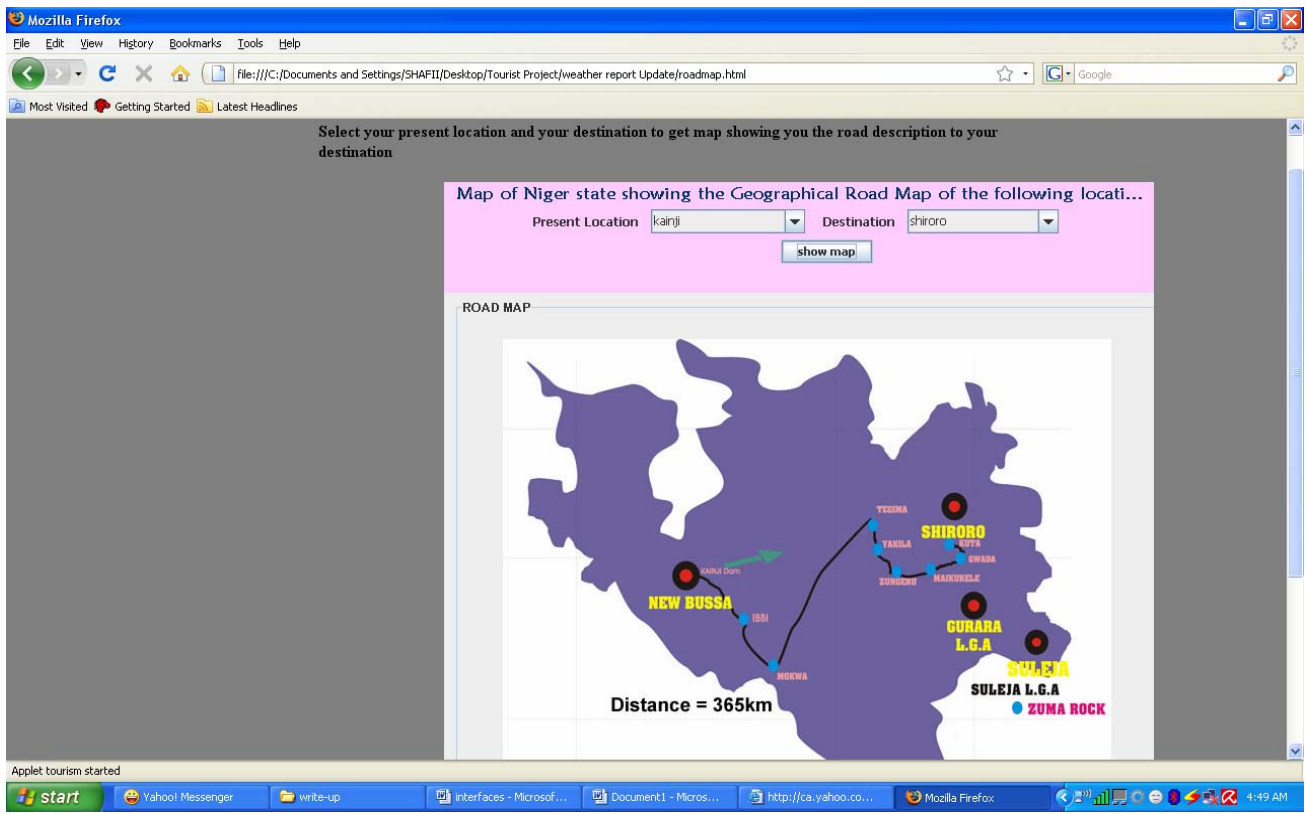

Figure 5: Road Map Navigation System for Tourists

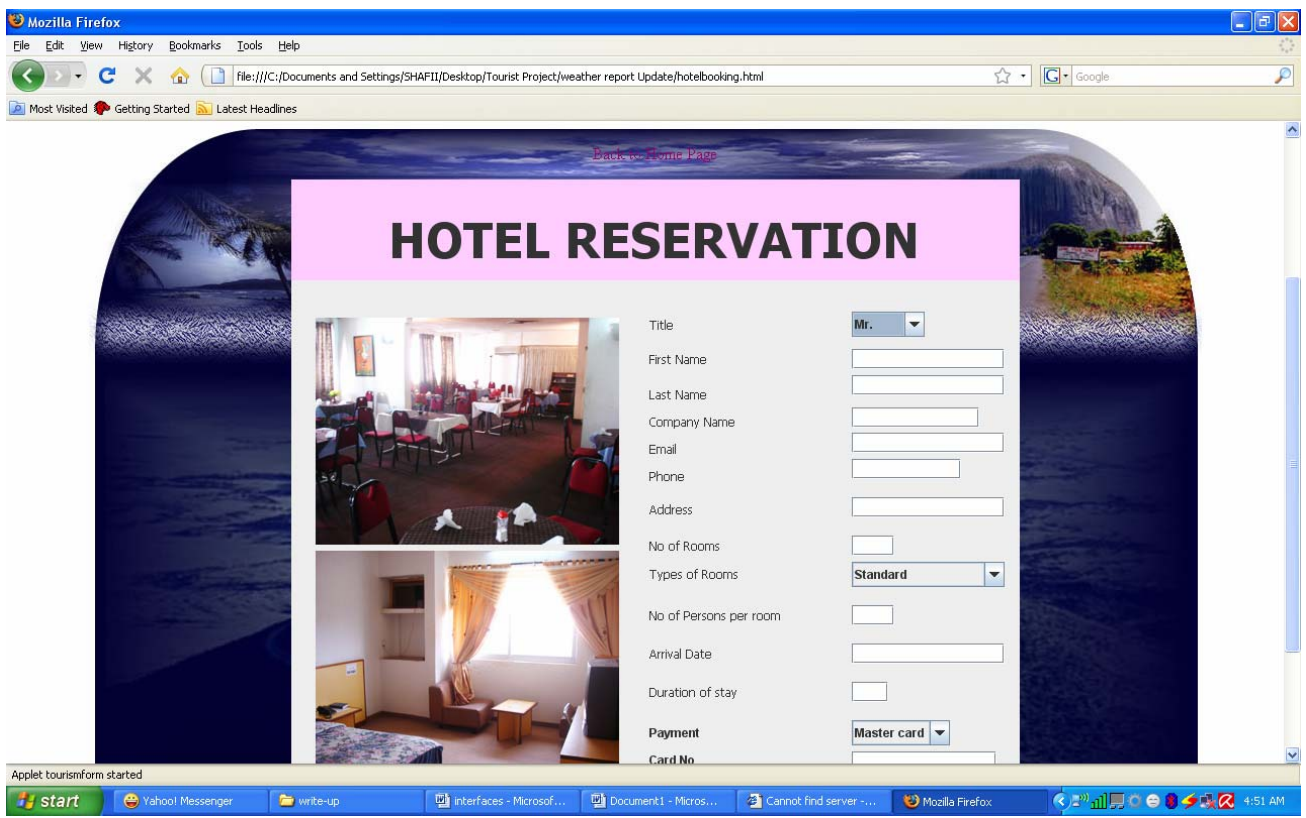

Figure 6: Hotel Reservation for Tourists

**System Testing**

The strategy for Web Application testing adopts the basic principles for all software testing and applies a strategy and tactics that have been recommended for object-oriented systems. The software was tested using three different Web browsers namely Internet Explorer 8, Mozilla Firefox version 3 and Opera 9.5. Two different operating systems are also used to test run the software, these are; Microsoft XP and Linux (Ubuntu). Two different MySQL server are also used to test the program, that is MySQL query browser and WAMP5 server. All the test results are very encouraging and successful, with very little errors.

**Conclusion**

The developments of information technology have a high influence on tourism development. Poor information results in inadequate analysis, which leads to misguided policies on tourism management. The current problem has many socio-economic, institutional and environmental aspects. An information system has the task to collect, analyze and process existing information. It is an active object, which deals with information and information processes. Maps are a natural means of indexing and presenting tourism related information. Travelers are using maps to navigate during their travels and for preparing their routes. Moreover, maps exploit the two dimensional capabilities of human vision and present the information in a compact and easy to read way. Through the utilisation of computer technology, new classes of operations based on adjacency, distance, proximity and route optimisation were made available to the final user in addition to more traditional multimedia data navigation and presentation functionality.